# Relics Theory of Dark Matter In quantum Evaporating Black Hole

Liao Liu[*]

Department of Physics, Beijing Normal University, Beijing, 100875, PRC

We introduce first a new road to quantize the Schwarzschild black hole. Then we find that the ground state or relic of the quantum evaporating black hole can be identified as dark matter.

PACS numbers：95.35.+d   04.70.Dy

Early in the thirties of the last century, astrophysicists began to know the existence of so-called missing mass or dark matter in our universe[1].Recent astronomical observation found doubtless about 22% of the matter of our universe are from dark matter[2]. Now the most important problem confronts us is what are the origin, candidates and properties of dark matter. Many authors tried to establish a theory of dark matter based on standard particles physics, but it seems no definite progresses obtained till now. So let us try a new road to study this mysterious problem. In the nineties of the last century, Carr et al, Barrow and Mac Gibbon conjectured that dark matter may be something consists from the relics or remnants of the evaporation black hole [3]. But let us emphasize that the existence of relics of an evaporating black hole in their theory is only a conjecture without sound support from theoretic physics. On the contrary, many physicists, e. g. Hawking et al maintained that nothing can be left due to the final violent explosion of evaporating black hole.

Now, however, in this short report, we shall point out though, there is no satisfied quantum gravity theory nowadays, we can still established a so-called quantum evaporating Schwarzschild black hole (QSBH) theory [4]. In this theory the SBH was first treated in Euclidean Kruskal section as a single periodic motion, then the famous Bohr-Sommerfeld quantum condition of motion

$$I_v = \oint pdq = 2\pi\hbar \cdot n \qquad (1)$$

was applied. Landau and Lifshitz derived the Sommerfeld quantum condition for the one-dimensional periodic motion of a particle by using the quasi-classical method and obtained [5]

$$\oint pdq = 2\pi\hbar(n+\frac{1}{2}) \qquad (2)$$

that is, for any cyclic motion the action variable $I_v$ of the system is quantized according to condition (2).

Now we try to apply this principle to the SBH. As is known, the Euclidean Kruskal section of SBH is a cyclic or single periodic system, whose metric reads

$$ds^2 = \left[\frac{32}{r}m^3 \exp(-\frac{r}{2m})\right](dT^2 + dR^2) + r^2 d\Omega_2^2, \quad (r > 2m) \qquad (3)$$

where

$$iT = (\frac{r}{2m} - 1)^{1/2} \exp(\frac{r}{4m}) \sin(\frac{\tau}{4m}) \qquad (4)$$

$$R = (\frac{r}{2m} - 1)^{1/2} \exp(\frac{r}{4m}) \cos(\frac{\tau}{4m}) \qquad (5)$$

Obviously, both $T$ and $R$ in the parametric form Eqs. (4) and (5) are the periodic function of $\tau$ with period $8\pi m$. We would like to emphasize that this peculiar property of the Euclidean-Kruskal metric is very important to reveal the thermodynamics of the SBH. We shall see in the following that it is also of key importance for the recognition of its quantum property.

From classical general relativity, we know that the area A of the event horizon of the SBH and its action I are, respectively,

$$A = 16\pi m^2 \quad (G = C = 1) \qquad (6)$$

$$I = 4\pi m^2 = \oint p dq - 8\pi m^2, \quad (G = C = \hbar = 1, T = 8\pi) \qquad (7)$$

Here we note that, for the vacuum Euclidean-Kruskal section, the volume integral of its action is zero, so the contribution to action comes only from the Gibbons-Hawking's surface term $4\pi m^2$.

It is noticed that there is an important relation between the action I and the action variable $I_v$ of the single periodic motion, i.e.

$$I = I_v - \oint H dt \qquad (8)$$

where $H$ is the Hamiltonian of the system. In the so-called (3+1) decomposition of any Einstein gravitational system, we have however two different kinds of Hamiltonians, i.e. the one $H(h_{ij}, \pi_{ij})$ without the Gibbons-Hawking surface term that satisfies the well known Hamilton constraint

$$H(h_{ij}, \pi_{ij}) = 0 \qquad (9)$$

and the other

$$H' = H + \alpha \qquad (10)$$

where

$$\alpha = \lim_{r \to \infty} \oint (\frac{\partial h_{ij}}{\partial x^j} - \frac{\partial \pi_{jj}}{\partial x^i}) ds^i \qquad (11)$$

is just the ADM mass m for asymptotic flat SBH. We recall that $h_{ij}$ is the induced metric of the three-dimensional space-like hypersurface and $\pi_{ij}$ is the canonical conjugated momentum of $h_{ij}$. The important thing in Eq. (10) we would like to emphasize is that the Hamiltonian $H'$ of any classical Einstein gravitational system with asymptotic flatness is always the ADM mass $m$ of the system. Thus, an important result for any Einstein single periodic gravitational system with boundary is

$$I = I_v - \oint H' dt = \oint pdq - m\oint dt = \oint pdq - mT \qquad (12)$$

From Eqs. (6) and (7) the variation $\Delta A$ of $A$ and $\Delta I$ of $I$ have the relation

$$\Delta A = 4\Delta I \quad (G = C = \hbar = 1) \qquad (13)$$

Now noticing $H' = m$ and $I = I_v - 8\pi m^2$ for the SBH, we can apply Sommerfeld's quantum conditions to obtain the spectrum of action $I$, the spectrum of area $A$ of event horizon and the spectrum of entropy $S$ of SBH as follows:

$$I = 4\pi m^2 = 2\pi\hbar(n+\frac{1}{2}) - 8\pi m^2 \qquad (14)$$

$$m^2 = \frac{1}{6}(n+\frac{1}{2})m_p^2 \qquad (15)$$

$$A = 16\pi m^2 = \frac{8\pi}{3}(n+\frac{1}{2})l_p^2 \qquad (16)$$

and

$$S = \frac{1}{4}A \qquad (17)$$

where

$$I = S = \frac{1}{4}A \qquad (18)$$

The minimum variation or quantum of the area of the event horizon $\delta A$, quantum of the entropy $\delta S$ and quantum of the mass $\delta m$ of SBH are, respectively,

$$\delta A = \frac{8\pi}{3}l_p^2 \qquad (19)$$

$$\delta S = \frac{2\pi}{3}k_B \qquad (20)$$

$$\delta m = \frac{1}{12}\frac{m_p^2}{m_T} \qquad (21)$$

where $l_p = (G\hbar C^{-3})^{1/2}$ is the Planck length, $l_p^2 = G\hbar C^{-3}$ is the Planck area, $m_p = (G^{-1}\hbar C)^{1/2}$ is the Planck mass and $m_T$ is the mass of SBH at temperature $T$.

It seems that our area quantum (19) only refers to the event horizon. We have no reason to infer that this is a general result to all areas of any surface. It is emphasized that our result of the quantum area (19) is different from the recent value of $4(ln3)l_p^2$ obtained by Dreyer [6]. Equation (21) is the total quantum mass-energy loss (QML) of an SBH via Hawking evaporation in temperature $T$. If the black hole mass $m(T)$ has a lowest limit $\frac{1}{2\sqrt{3}}m_p$, then the QML has an upper limit $\Delta m = \frac{1}{2\sqrt{3}}m_p$

From Eqs. (21), (22) and (23) we see $n = 0$ should correspond to the ground state of SBH. It is easy to show the ground state mass $m_G$ of the SBH is

$$m_G = \frac{1}{2\sqrt{3}}m_p \qquad (22)$$

It seems that there is no way to decrease the mass of an SBH under its ground state mass $m_G$. Therefore even Hawking evaporation will cease as a QSBH approaches its ground state; in other words, the QSBH will transit into something other than black hole. The important thing is that there is a remnant left for any evaporating black hole. The black hole cannot annihilate away totally due to Hawking evaporation.

Therefore, several interest results are:

The first is an incoming pure vacuum state can never become an outgoing mixed state! No loss of quantum coherence will result in our quasi-classical scheme of quantum gravity. In other words, the long unsolved information puzzle in quasi-classical black hole physics may have a solution now! This is a very interesting result.

The second is that the ground state (or relics, remnants) of QSBH may be really called a dark star or a "dead black hole" without any kind of radiation the existence of them can only be detected by their gravitational action on other stars. So a very natural and more reasonable conjecture is that the principal constituent of the dark matter may come from the above mentioned dark stars.

So quantum mechanics forbids the total annihilation of black hole through evaporation, as a result, relics or remnant should accompany with black hole. It is easy to see that the temperature $T_D$ of relic is an universal constant

$$T_D = \frac{1}{2\sqrt{3}}T_P \quad (T_D = (G^{-1}\hbar C^5 k^{-2})^{1/2}) \qquad (23)$$

Though it is a very high Planck scale temperature, it still can't give out any kind of radiation, in this sense all the relics are dark. This is a very surprised property. Henceforth the origin of dark matter is just the relic or ground state of quantum evaporation black hole! No other things can be looked upon as the origin of dark matter!

If dark matter is really the ground state of evaporating black hole we can demand that the ground state should have no hair that means e. g. we can't discriminate a dark star of axion to that of neutrolinos or the number of parameters characterize a dark star can't greater than mass, charge and angular momentum

* Email:   liuliao1928@yahoo.com.cn